\documentclass[twocolumn,english,aps,showkeys,showpacs]{revtex4}
\usepackage[T1]{fontenc}
\usepackage[latin9]{inputenc}
\usepackage{graphicx}
\usepackage{amssymb}

\makeatletter
\@ifundefined{textcolor}{}
{%
 \definecolor{BLACK}{gray}{0}
 \definecolor{WHITE}{gray}{1}
 \definecolor{RED}{rgb}{1,0,0}
 \definecolor{GREEN}{rgb}{0,1,0}
 \definecolor{BLUE}{rgb}{0,0,1}
 \definecolor{CYAN}{cmyk}{1,0,0,0}
 \definecolor{MAGENTA}{cmyk}{0,1,0,0}
 \definecolor{YELLOW}{cmyk}{0,0,1,0}
 }

\makeatother

\makeatother

\usepackage{babel}

\makeatother

\usepackage{babel}

\makeatother

\usepackage{babel}

\usepackage{babel}

\makeatother

\usepackage{babel}

\begin{document}

\title{Dipolar interactions in arrays of ferromagnetic nanowires: a micromagnetic
study}

\author{Fatih Zighem, Thomas Maurer, Frédéric Ott and Grégory Chaboussant}

\affiliation{Laboratoire Léon Brillouin, IRAMIS, CEA-CNRS UMR 12, CE-Saclay, 91191
Gif sur Yvette, France}
\begin{abstract}
We explore the behavior of periodic arrays of magnetic nanowires by
micromagnetic simulations using the Nmag modeling package. A large
number of modeling studies on such arrays of nanowires have been performed
using finite size models. We show that these finite size micromagnetic
descriptions can only be used in specific situations. We perform a
systematic study of more or less dense 1D and 2D arrays of nanowires
using either finite size or infinite size models and we show that
finite size models fail to capture some of the features of real infinite
systems. We show that the mean field model scaled to the system porosity
is valid. This work can be used as a basis to the extension of micromagnetic
calculations of the magnetization dynamics in arrays of nanowires.



\end{abstract}
\maketitle

\section{Introduction}

During the past decade, fast progress has been made in the synthesis
of nano-objects. In particular, efforts have been made in the synthesis
of magnetic nanowires. Several routes have been developed for the
preparation of such objects. The most studied consists in the electrochemical
deposition using porous polycarbonate or anodic alumina (Al$_{2}$O$_{3}$)
membranes as templates \cite{sellmyer2001,hurst2006,whitney,pan,ansermet,zeng,han,zhang,vazquez}.
The static magnetic properties have been intensively studied by magnetometry
experiments \cite{sellmyer2001,hurst2006,whitney,pan,ansermet,zeng,han,zhang,vazquez}
while the dynamic magnetic properties have been probed by Ferromagnetic
Resonance (FMR) \cite{demand2002,Sklyuyev,kartopu,darques} and Brillouin
Light Scattering (BLS) \cite{nguyen2006,wang2006,andrei} experiments.
Among the various issues at stake for a comprehensive understanding
of these materials is the influence of long-range dipolar interactions
between nanowires. In fact, as shown in previous studies, the dipolar
coupling between neighboring wires can considerably reduce the coercive
field of an assembly of wires \cite{velazquez1999,sampaio2000}. In
addition, the magnetic moments distribution inside a wire can be strongly
affected by stray fields originating from neighboring wires, and the
usual coherent rotation approximation is then not justified \cite{jaccard2000}.
Another manifestation of dipolar interactions concerns the magnetic
excitations (spin waves) which were shown to be strongly dependent
on the demagnetizing field and on surfaces and interfaces boundaries
\cite{nguyen2006,zighem2007}. A rigorous analysis of the static and
dynamic measurements requires then to take into account these dipolar
interactions. In some cases, dipolar interactions are taken into account
by considering nanowires as dipoles \cite{velazquez1999,piccin} but
this approximation is only justified when the distribution of magnetic
moments inside the nanowire is uniform, allowing for an analytical
treatment of the magnetic interaction between neighboring wires\cite{yelon}.\\

The scope of this study is to quantify the dipolar interactions in
various geometries of nanowires arrays (see Figure 1). Our micromagnetic
calculations take explicitly into account the dipolar interactions
between nanowires, and we focus on the static magnetic properties
via the calculation of magnetization curves measured in different
directions between the nanowire's axis and the applied magnetic field
direction. The paper is structured as follows: the magnetic parameters
of the studied systems are presented in section II. Section III and
IV are devoted to the study of the magnetostatic interactions in linear
rows of nanowires and in hexagonal arrays of nanowires, respectively.

\section{Magnetic parameters and geometry of the nanowires}

Throughout this communication, the basic building brick which is considered
is a Co magnetic nanowire of length ($\ell$) 100 nm and diameter
($d$) 10 nm. These dimensions are typical of real materials \cite{andrei,thurn,vidal}.
The diameter of the wire has been chosen so that a coherent rotation
is expected. The diameter for coherent magnetic rotation in Co spheres
is 15 nm. The length of the wire has been chosen to be 10 times the
diameter so that the shape anisotropy is very close to that of an
infinite cylinder. The demagnetizing coefficient for an ellipsoid
with an aspect ratio of 10 is 0.017 which is very close to 0. The
limitation to a length of 100 nm allows to perform more complex micromagnetic
simulations than would be possible with longer wires while keeping
the key property of a strong shape anisotropy.\\

The magnetic parameters used correspond to typical value for hcp cobalt
thin films \cite{tannewald}: magnetization saturation $M_{S}=1.4\times10^{6}$
A.m$^{-1}$, exchange stiffness $A=1.2\times10^{-11}$ J.m$^{-1}$,
the magnetocrystalline anisotropy is neglected. The exchange length,
defined as $\ell_{ex}=\sqrt{2A/\mu_{0}M_{S}^{2}}$, is around $3.2$
nm. We have used the Nmag micromagnetic modelling package \cite{fishbacher2007},
which is a finite element code, in order to solve the Landau-Lifshtiz-Gilbert
equation. In our simulations, the damping (Gilbert) constant $\alpha$
was set to $0.5$ in order to minimize the computing time and because
we are only interested in the static magnetization configurations
\cite{fishbacher2007}. The nanowires were meshed with Netgen \cite{netgen}
so that the distance between two nodes is of the order of 2 nm, being
smaller than the exchange length ($\ell_{ex}$). We have checked that
the results obtained with a finer mesh ($1$ nm between two nodes)
are identical. Our convention is that the wires axis is $Ox$ direction
(see Figure 1a). The magnetization reversal of the different systems
has been studied through magnetization curves calculations along the
three directions ($Ox,Oy$ and $Oz$). The magnetization curves are
investigated by applying external magnetic fields $B$ ranging from
$-1800$ mT to $+1800$ mT and back to $-1800$ mT, the field is incremented
in steps of $12$ mT. Note that $m_{i}$ corresponds to the projected
reduced magnetization ($M_{i}/M_{S}$ ) along $i$ ($x,y$ or $z$).\\

For a field applied along the wire axis in the case of an isolated
nanowire, a square hysteresis loop is obtained which indicates that
the nanowire switches between two homogeneously magnetized states
\cite{ott2009,maurer2009}. The coercive field
$\mu_{0}H_{C}\simeq470$ mT to be compared with
$\mu_{0}H_{C}\simeq827$ mT in the case of an ellipsoid with the same
aspect ratio \cite{osborn}. For a field applied perpendicular to the
wire axis (along $Oy$ or $Oz$: hard axis directions), the saturation
field is $\mu_{0}H_{S}\simeq800$ mT which is very close to that of
an ellipsoid with the same aspect ratio $\mu_{0}H_{S}\simeq827$ mT
(see supplementary Figure S1) .

\begin{figure}[h]
\includegraphics[bb=85bp 185bp 550bp 570bp,clip,width=8cm]{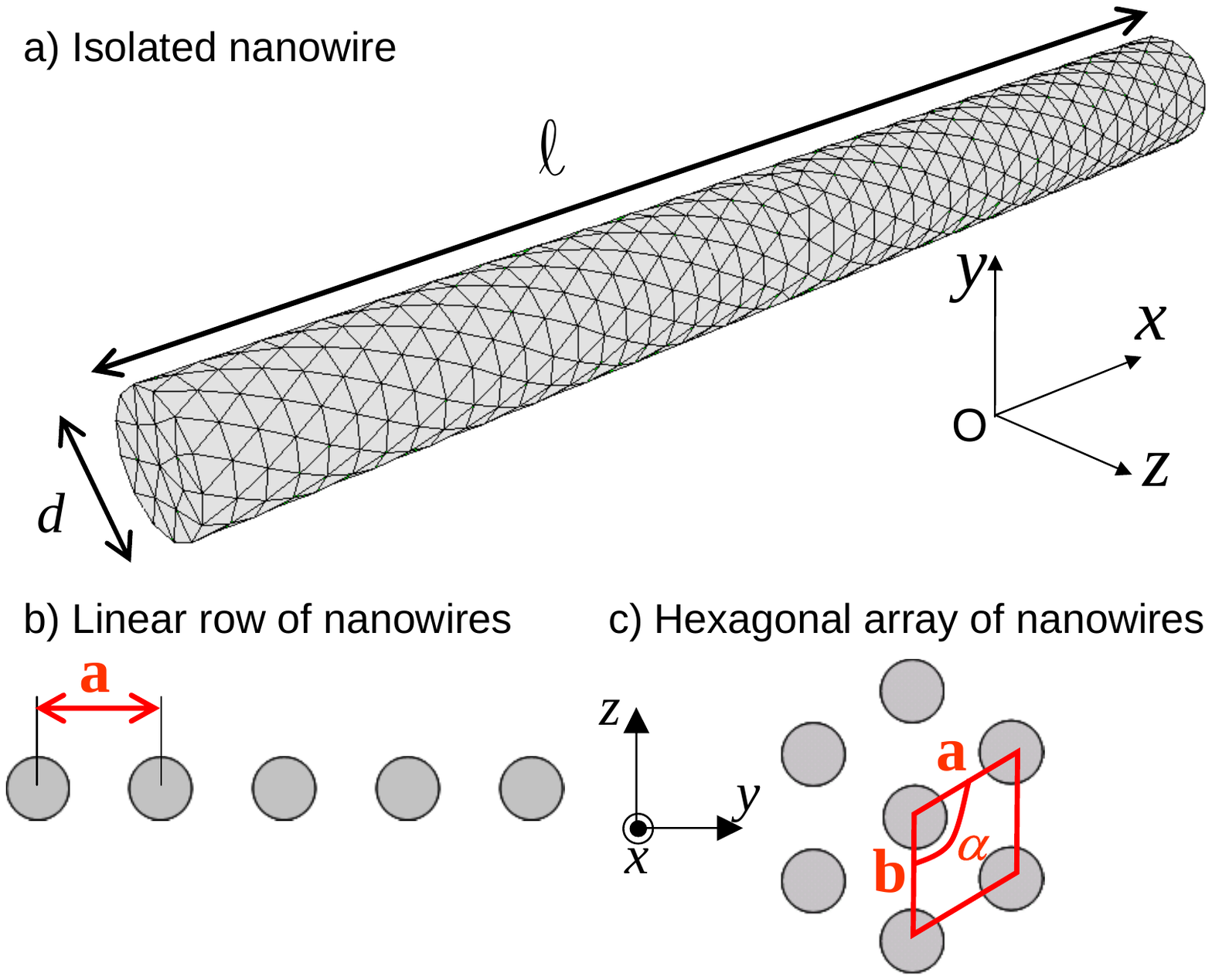}
\label{mesh-1}

\caption{a) Typical mesh used for the micromagnetic calculations
($\ell=100$ nm and $d=10$ nm). The mesh is composed of 700 nodes and
the distance between two nodes is around $2$ nm. The wire axis is
along the $Ox$ direction. b) Top view (($yOz$) plane) of a linear
row of 5 nanowires along $Oy$. c) Top view of an hexagonal array of
7 nanowires ($a=b$ and $\alpha=120^{\circ}$) .}

\end{figure}

\section{Linear rows of nanowires}

\subsection{Finite rows}

In order to address the influence of dipolar interactions, we have
first examined the case of a finite number of nanowires ($N=3$ to
13), ordered parallel to each other\textbf{ }along the $Oy$ direction
(see Figure 1b). The distance between the wires centers ($a$) was
varied from 15 to 60 nm, that is from $1.5d$ to $6d$. For an inter-wire
distance larger than 50 nm, we did not find any significant difference
in the magnetization curves from the simple case of an isolated nanowire.
\\

The stable remanent states in these row arrangements depend on both
the amplitude and direction of the magnetic field initially applied.
The remanent state obtained after applying a saturating field along
$Ox$ corresponds to an alignment of all the magnetic moments along
$Ox$ even for small $a$-values. On the other hand, the remanent
state after applying a saturating field along $Oy$ or $Oz$ depends
on both the distance $a$ and the number of interacting nanowires.
The general behavior is that the magnetic moments inside the nanowires
are quasi-homogeneously magnetized and only the total magnetization
direction ($+Ox$ or $-Ox$) changes from a nanowire to the next.
However, the system does not necessarily adopt a perfect {}``antiferromagnetic''
($\cdots$up-down-up-down$\cdots$) configuration.\\

\begin{figure}[!h]
\includegraphics[bb=25bp 195bp 390bp 587bp,clip,width=8cm]{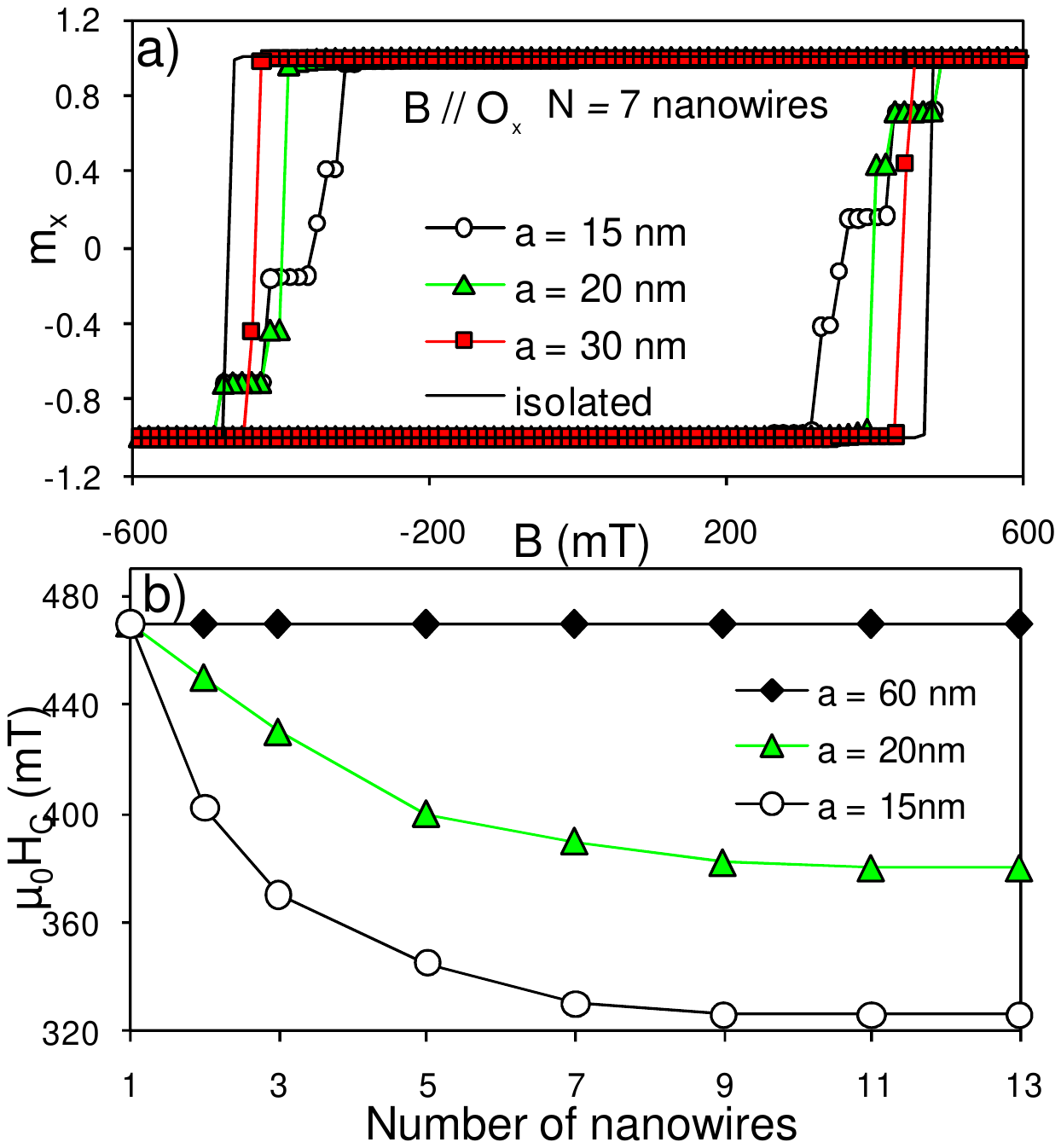}\label{Hc_Chains}

\caption{a) Magnetization curves calculated with $B\parallel Ox$ for different
inter-wire distances in a row formed by $N=7$ nanowires. The solid
black line corresponds to the hysteresis cycle of an isolated nanowire.
b) Evolution of the coercive field of the central nanowire as a function
of the number of nanowires. Calculated values for three different
inter-wire distances ($a=15,20,60$ nm) are shown. Lines are guided
for the eyes.}

\end{figure}

Figure 2a presents typical magnetization curves obtained for a 7 nanowires
row as a function of the inter-wire distance and a field applied along
Ox. The magnetization curves present plateaus which correspond to
the switching of a single wire of the row. This was experimentally
put into evidence in magnetic microwires by Sampaio et al . \cite{sampaio2000}.
However, in our simulations we observe that the widths of the plateaus
tend to increase when $a$ decreases and that the last plateaus (next
to the last and last wire to switch) are more extended than the first
plateaus (see Figure 2a). This is qualitatively different from the
behavior experimentally observed in \cite{sampaio2000}. We attribute
this to the fact that in the experimental measurements, $d\ll a$
so that dipolar effects were limited. On the other hand, in our calculations,
$d\sim a$, so that dipolar effects play a key role: the energy barrier
for the last wire switch is thus greatly enhanced because of dipolar
fields from the other wires which opposes the last wire switch. .

Figure 2b presents the switching field of the central nanowire for
different types of rows. In all cases, the central nanowire switches
first due to a maximum intensity of the dipolar field created by the
surrounding nanowires. Figure2 illustrates 2 points: (i) there is
an inter-wire distance ($a>60$ nm) for which the dipolar coupling
can be neglected. (ii) when dipolar coupling between wires is significant
($a<60$ nm), it is nevertheless only necessary to take into account
a rather small number of neighbours (3 on each side) to converge towards
the infinite case. This shows that demagnetizing field effects are
small in the case of 1D chains.

\begin{figure}[!h]
\includegraphics[bb=25bp 250bp 390bp 585bp,clip,width=8.5cm]{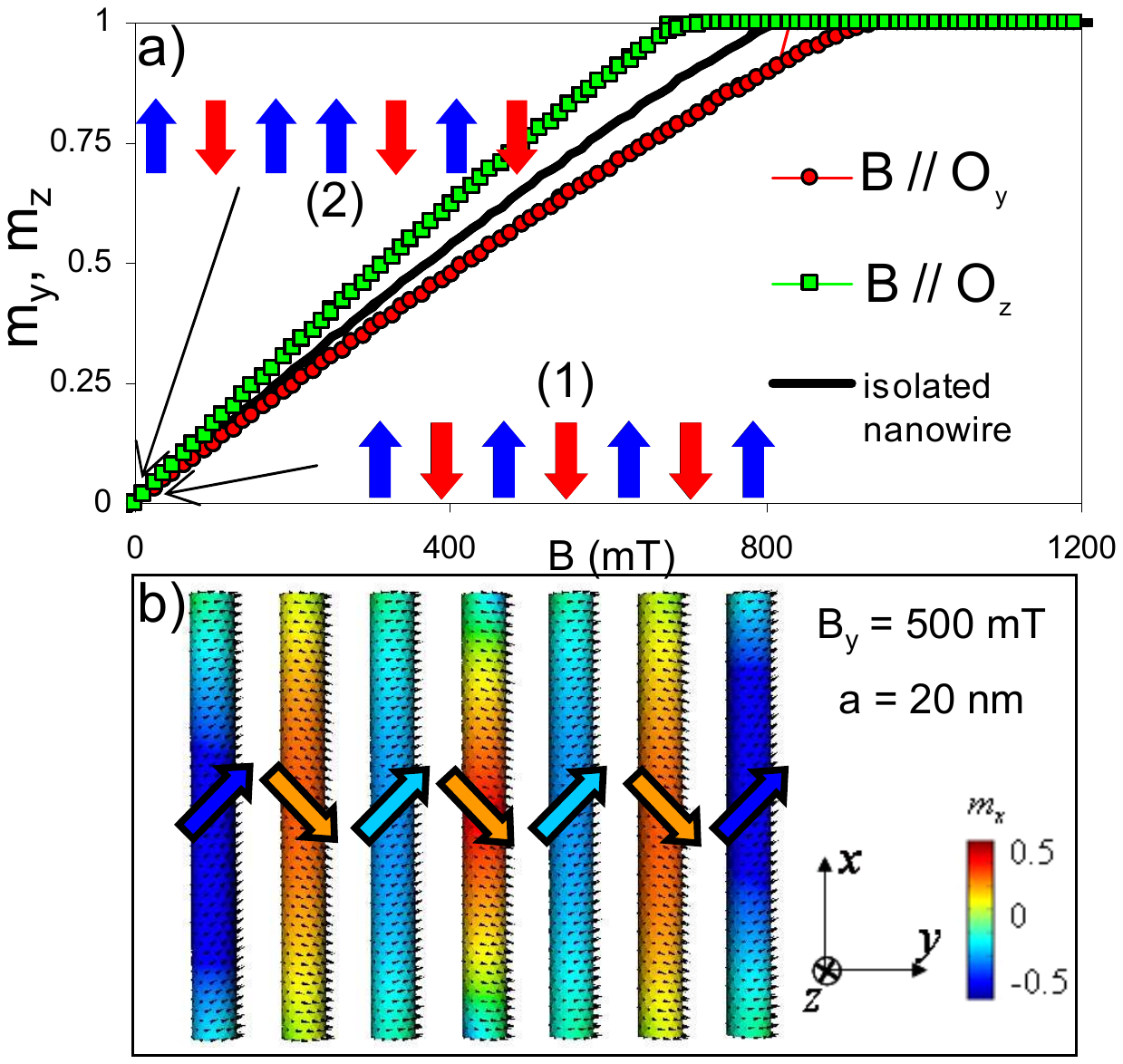}\label{Curves_7nanowires}

\caption{a) Magnetization curves calculated along $y$ (triangles) and $z$
(circles) for a row of $N=7$ nanowires ($a=20$ nm). The solid line
is the magnetization curve along the hard axis of an isolated nanowire.
We have also represented a sketch of the remanent state obtained after
applying $B$ along $Oy$ (1) and along $Oz$ (2). b) Magnetic moments
distribution when the system is subjected to a transverse (along $Oy$)
magnetic field of 500 mT. }

\end{figure}

Figure 3a presents the magnetization curves for a field $\vec{B}$
applied along $Oy$ and $Oz$ on a row of $N=7$ nanowires. The saturation
along $Oz$ is reached for an applied field larger than along $Oy$.
Indeed, when the nanowires are close (i. e. $a\rightarrow10$ nm)
and for a large number of nanowires, the system can be viewed as an
infinite strip along $Oy$, which defines a second easy axis (in addition
to the one along $Ox$ defined by the shape anisotropy of the nanowires).
In these conditions, the $Oz$ direction can be defined as a hard
axis and a larger applied field will be necessary to reach the magnetization
saturation in this direction (see Figure 4a for an infinite row of
nanowires). The dipolar field created by a given wire strongly depends
on the magnetic moments distribution inside the wire; thus, when the
magnetic moments begin to orient along the $Oz$ direction (with $B\parallel Oz$),
the stray fields created on its neighbors will point in the $-Oz$
direction (in first approximation) and will not favor alignment along
$Oz$. On the contrary, when the magnetic moments begin to orient
along the $Oy$ direction (with $B\parallel Oy$), the stray fields
created on its neighbors favor an alignment along $Oy$ (see Figure
3b). Note that, as illustrated in Figure 3b, non uniform magnetic
moments distributions appear inside the wires under applied magnetic
field. This effect could be a source of localized spin waves modes,
as proposed in Ref. \cite{jorzick}.

\subsection{Infinite rows}

We have used the hybrid {}``finite element/boundary element'' method
\cite{nmag-hybrid} available in the Nmag package to model an infinite
row of nanowires. A particular importance is given to the choice of
the elementary meshed cell as the results can strongly depend on it.
The elementary cell is composed of 10 wires in order to have a large
cell and a reasonable computing time. In this method, to take into
account the long-range dipolar interaction between the nanowires,
virtual copies of the elementary cell are created \cite{fishbacher2007,nmag-hybrid}.
40 virtual copies (20 to the left and 20 to the right of the meshed
cell) have been used in the following simulations.\\

\begin{figure}[!h]
\includegraphics[bb=30bp 210bp 390bp 585bp,clip,width=8.5cm]{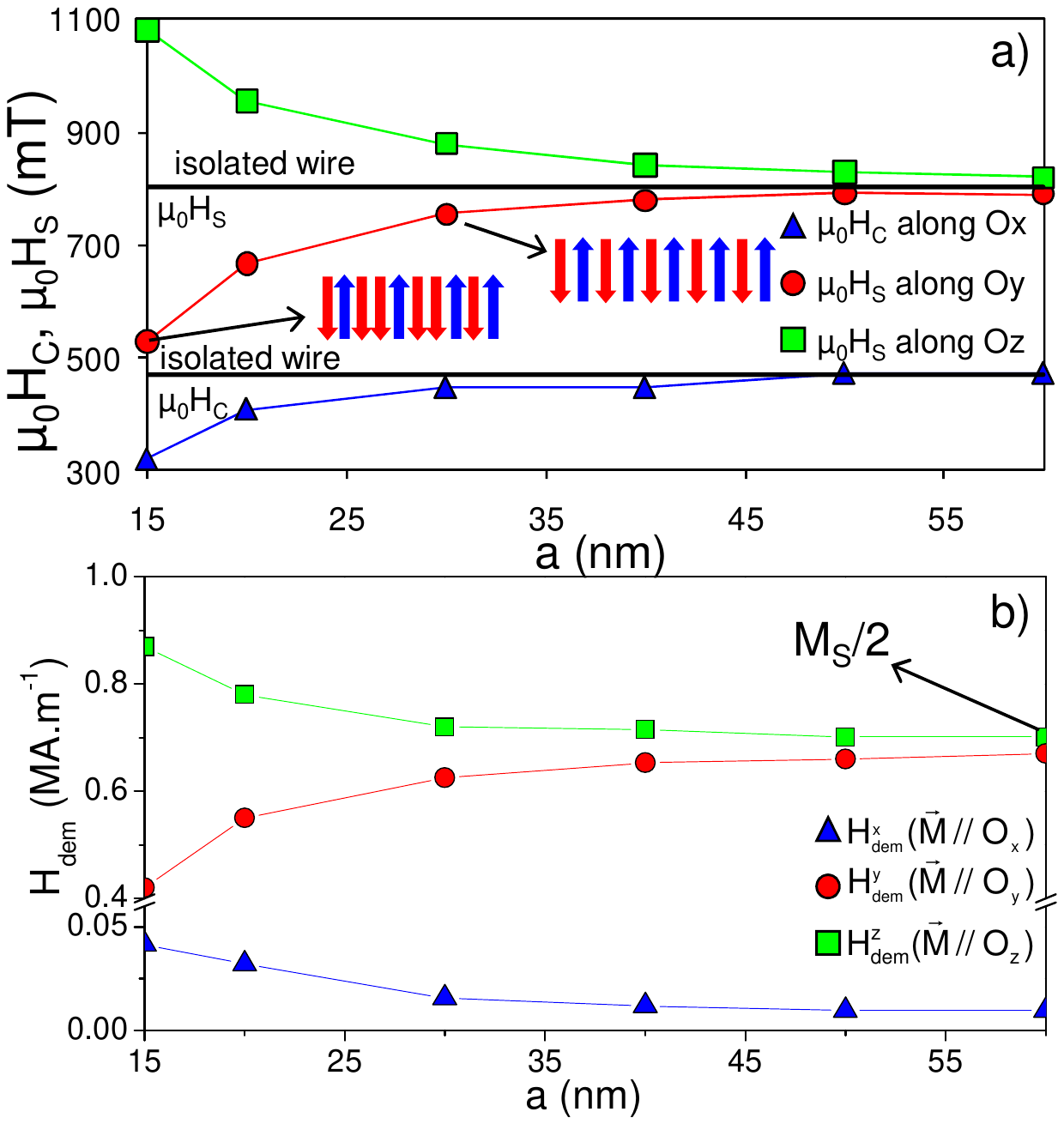}\label{Curves_Chain}

\caption{a) Coercive (triangles) and saturation (circles and squares) fields
obtained from calculated magnetization curves along $Ox$, $Oy$ and
$Oz$ direction, respectively. Sketches represent zero field configurations
after applying a saturating field along $y$. Lines are guided for
the eyes. b) $x$- (triangles), $y$- (circles) and $z$- (squares)
components of the demagnetizing field calculated in an infinite row
respectively saturated along $Ox$, $Oy$ and $Oz$ as a function
of the inter-wire distance $a$.}

\end{figure}

When $B\parallel Ox$, the zero field configuration corresponds to
a parallel alignment of the magnetic moments along $Ox$. The wires
are subjected to the same dipolar field created by the neighboring
wires and switch all for the same applied field. The net result is
a reinforcement and an homogeneization of the demagnetizing field
which explains why the magnetization is reversed for a smaller applied
field compared to the isolated wire situation. The coercive fields
$\mu_{0}H_{C}$ in infinite rows (see Figure 4a) are then very close
to the ones obtained for the central nanowire of rows composed by
9 or more nanowires (see Figure 2b).\\

When $B\parallel Oy$ or $Oz$, the zero field state corresponds to
magnetized wires randomly aligned along $+Ox$ or $-Ox$ and we find
that it strongly depends not only on the distance $a$ but also in
the choice of the elementary cell. Figure 4a presents the saturation
fields $\mu_{0}H_{S}$ obtained from the magnetization curves calculated
along $Oy$ and $Oz$. We found that $\mu_{0}H_{S}$ (measured along
$Oy$ or $Oz$) is not affected if a larger elementary cell is chosen
despite the fact that different remanent state are obtained. This
indicates that defects in the perfect {}``antiferromagnetic'' remanent
state ($\cdots$up-down-up-down$\cdots$) have pratically no influence
in the magnetization reversal. Finally, as expected, for $a$ larger
than 50 nm, the dipolar interactions are negligible and the macroscopic
($\mu_{0}H_{S}$ and $\mu_{0}H_{C}$) values measured are close to
the ones obtained for isolated wires.\\

In addition, we have represents (see Figure 4b) the $x$-, $y$- and
$z$-components of the demagnetizing field calculated in the infinite
row respectively saturated along $Ox$, $Oy$ and $Oz$ as a function
of the inter-wire distance $a$. As explained previoulsy, when $a\rightarrow10$
nm, the system can be viewed as an infinite stripe along $Oy$. This
is confirmed by the $y$- (resp. $z$-) component of the demagnetizing
field which strongly decreases (resp. increases) when $a$ decreases
and explained why a smaller (resp. larger) field is necessary to saturate
the system along $Oy$ (resp. $Oz$) (see Figure 4a). Moreover, the
$x$-component of the demagnetizing slowly decreases with $a$; it
confirms that the easy axis along the $Ox$ direction is less pronounced
for small inter-wire distance. \\
\\
\\
\\

To summarize, these results show that some aspects of the static magnetic
properties of an infinite row of nanowires could be analyzed by using
a finite but sufficiently large number of interacting nanowires.

\section{Hexagonal arrays of nanowires}

In this section, we consider hexagonal arrays of Co nanowires. The
choice of this arrangement is motivated by the fact that synthesis
routes, such as electrodeposition, lead to the fabrication of well
defined hexagonal arrays of nanowires with excellent control of their
diameter ($d$), length ($\ell$) and inter-wire distance ($a$) \cite{thurn,vidal}.
For large inter-wire distances, the magnetic properties of these systems
are easily analyzed as one can neglect the dipolar couplings between
the nanowires. For small inter-wire distances (dense array of nanowires),
the dipolar coupling between the nanowires must be taken into account
in the static and dynamic magnetic properties. Until now, the magnetic
behavior of such systems are generally analyzed by using a finite
number of interacting nanowires \cite{hertel,hertel2002} or by a
phenomenological treatment of the dipolar coupling \cite{nielsch,delatorre,piraux}.
We propose here to compare finite and infinite hexagonal arrays of
nanowires.

\subsection{Finite size arrays}

Three different arrays are considered. They are defined by the number
of interacting nanowires: 7 (see Figure 1c), 16 ($4\times4$) and
30 ($6\times5$). The inter-wire distance ($a$) is varied from 15
to 40 nm, that is from $1.5d$ to $4d$. The length ($\ell$) and
diameter ($d$) of the nanowires are 100 and 10 nm, respectively.\\

The remanent state derived from magnetization curves calculated along
$Ox$ (see Figures 6a) corresponds to an alignment of the magnetic
moments along $x$. The magnetization reversal takes place in several
steps (see Figure 5a-b) defined by the magnetization switching of
one or more nanowires between $+Ox$ to $-Ox$. These steps are clearly
revealed by several magnetization plateaux in the hysteresis cycles.
Just before an individual nanowire switches, the $y$- and $z$-components
of the magnetic moments located at the tips significantly increase
and act as nucleation points for the magnetization reversal process.
Moreover, when the nanowires have almost completely switched, a plateau
of several mT can appear in the hysteresis cycle ($a=15$ nm and $B<-400$
mT in Figure 5b). This plateau is due to the finite size of the system:
the dipolar field created by the switched nanowires hinders the switch
of the last nanowires.\\

\begin{figure}[!h]
\includegraphics[bb=35bp 190bp 500bp 595bp,clip,width=8.5cm]{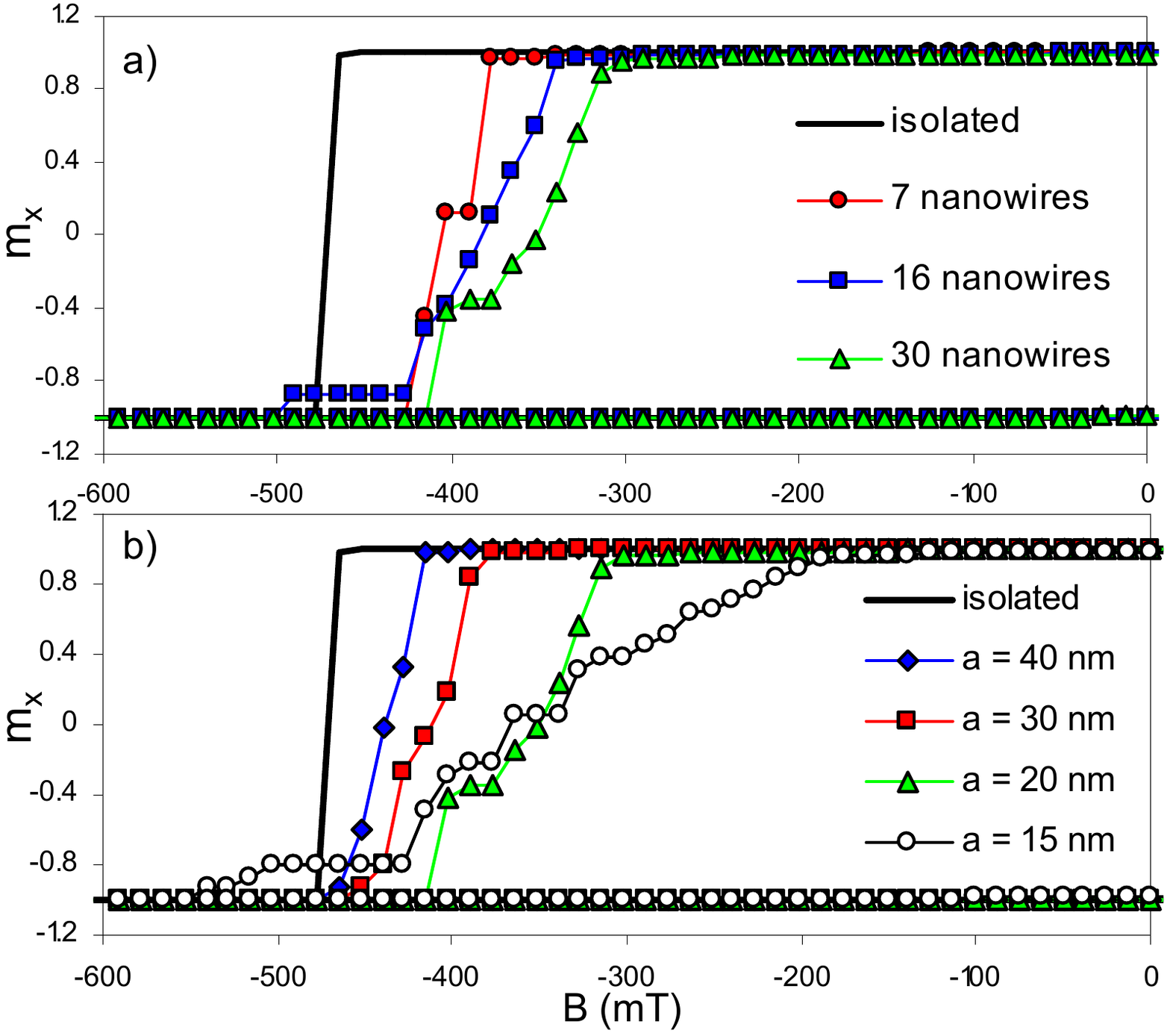}\label{Curves_Finite_Hexagonal}

\caption{a) Hysteresis curves for a field applied along $Ox$ on hexagonal
arrays of different sizes with $a=20$ nm. b) Hysteresis cycles calculated
for an array of 30 nanowires for different inter-wire distances. The
solid black line is the cycle obtained for an isolated nanowire.}

\end{figure}

\begin{figure}[!h]
\includegraphics[bb=25bp 135bp 510bp 580bp,clip,width=8.5cm]{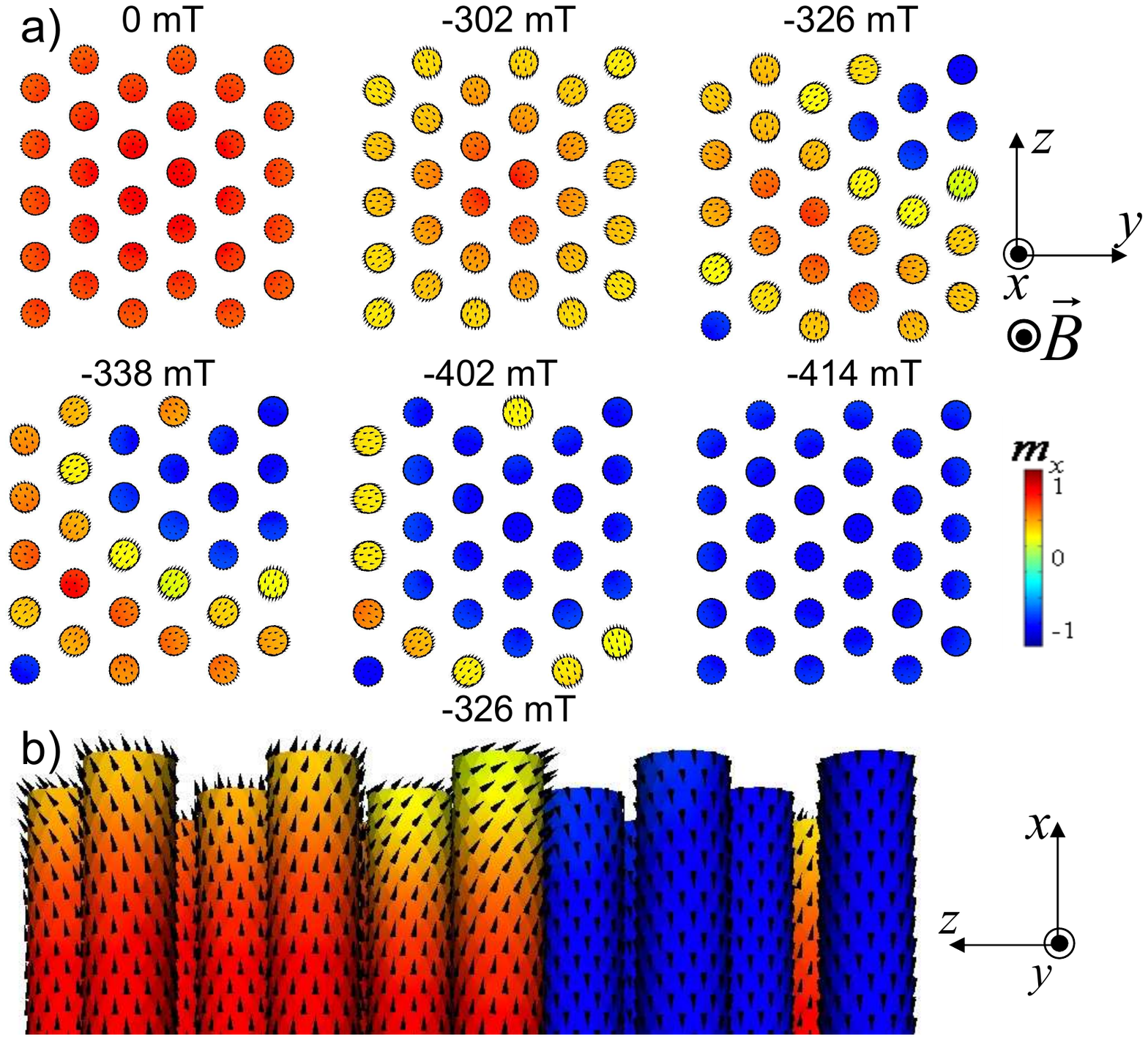}

\caption{Array of 30 interacting nanowires ($a=20$ nm): a) top view of the
magnetic moments distribution as a function of $B_{x}$. At zero applied
field, the magnetization is quasi-homogeneous and aligned along $Ox$;
when increasing $B_{x}$, the nanowires switch one after the other
and the system is saturated along $-Ox$ for an applied field larger
than $-415$ mT. Colors encode $m_{x}$. b) Side view of the system
($B_{x}=-326$ mT) showing the increased of the $m_{y}$ and $m_{z}$
components before the magnetization reversal.}

\end{figure}

Figure 6a shows a top view of the magnetic moments distribution obtained
for an array of 30 nanowires as a function of $B_{x}$. These micromagnetic
configurations have been extracted from the hysteresis cycle (with
$a=20$ nm). It can be observed that just before the magnetization
reversal ($B_{x}=-302$ mT), $m_{x}$ is close to zero at the tips
(see Figure 6b). Moreover, the reversal of the system behaves similar
to the domain wall propagation in thin films rather than a random
switching of the wires. When increasing the number of interacting
nanowires (see Figure 5a) or reducing the period of the array (see
Figure 5b), the hysteresis cycles are less square due to the dipolar
coupling between the objects; the coercive field is then reduced by
$25\%$ for 30 nanowires (with $a=20$ nm) compared to the coercive
field obtained for isolated nanowires.\\

\begin{figure}[!h]
\includegraphics[bb=50bp 120bp 410bp 550bp,clip,width=8.5cm]{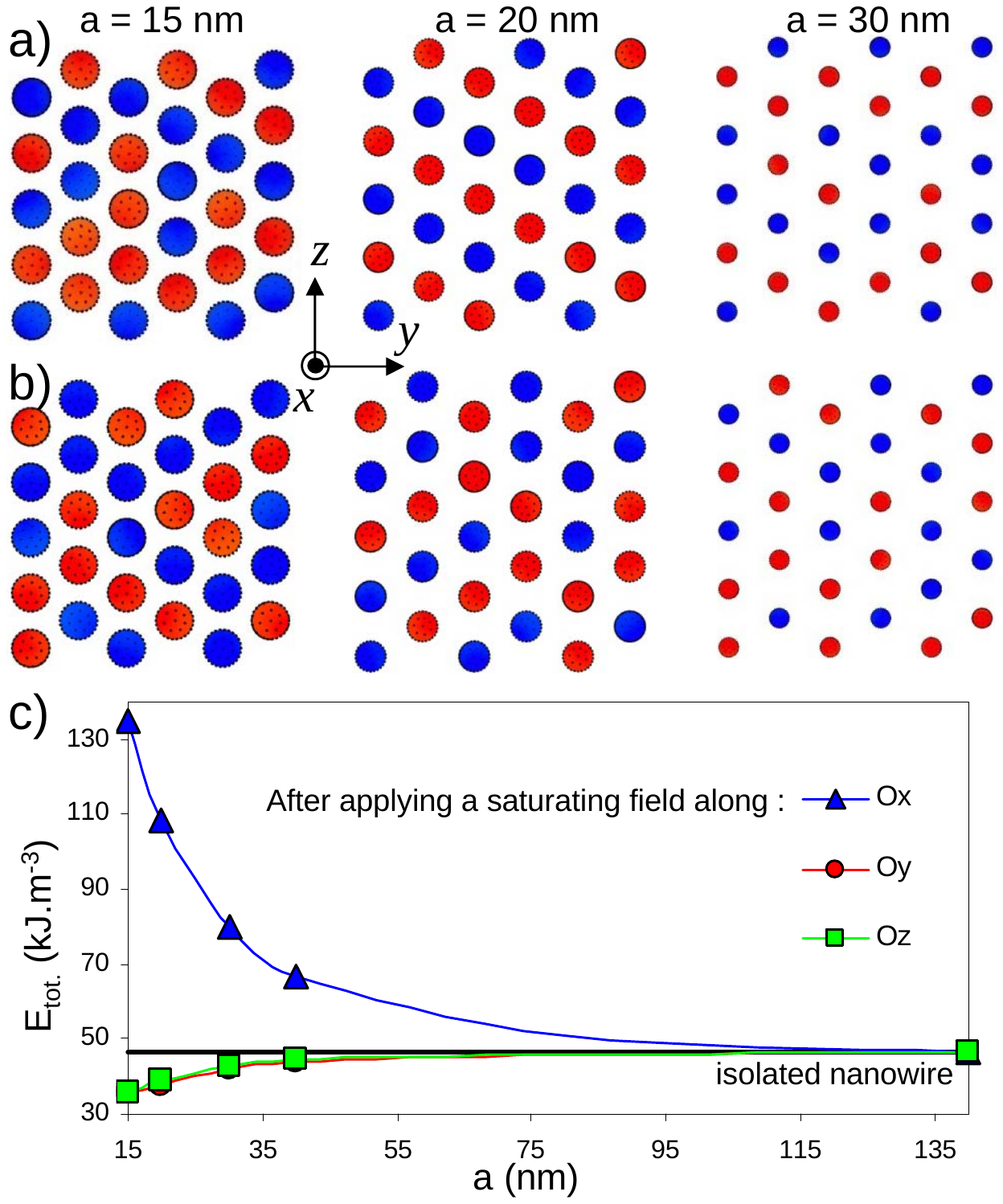}

\caption{Array of 30 interacting nanowires. a) and b) Remanent states after
applying a saturating field along $Oy$ (a) and $Oz$ (b) for different
inter-wire distances. b) Total energy at remanence after applying
a magnetic field along $x$, $y$ and $z$ as function of the inter-wire
distance.}

\end{figure}

The remanent state presented just above (quasi-homogeneously magnetized
nanowires) is not the only stable state. Other remanent micromagnetic
configurations appear when applying a saturating magnetic field along
$Oy$ or $Oz$ (see Figure 7). They correspond to a random orientation
$+Ox$ or $-Ox$ of the nanowire's magnetizations, apart for the case
($a=20$ nm; $B\parallel Oy$), where an antiferromagnetic order appears.
These remanent states are more stable than the one presented in Figure
8 because they minimize the demagnetizing field inside the nanowires
more efficiently. Figure 7c shows the total energy $E_{tot}$ (kJ.m$^{-3}$)
of the different remanent states. It shows that the states obtained
after applying a saturating field along $Oy$ or $Oz$ have a lower
total energy than the one obtained in absence of dipolar coupling.
$E_{tot}$ is large when all the nanowires are fully magnetized in
the same direction ($B_{sat}\parallel Ox$) and diminishes when $a$
increases (i. e. when the dipolar coupling weakens). For random wire
orientation ($B_{sat}\parallel Oy$ or $B_{sat}\parallel Oz$), the
magnetic energy decreases when $a$ decreases. This is due to the
fact that flux closure is more efficient for the random orientation
of the nanowires. \\

The magnetization curves calculated along $Oy$ and $Oz$ are identical
contrary to the row case (Section III). The applied field necessary
to ensure magnetization saturation along $y$ or $z$ is around 750
mT (for $a$ in the range \{15-40 nm\}), and it depends weakly on
the number of interacting nanowires $N$ (in the range $\{7-30\}$).
Moreover, inside a nanowire, the magnetization reversal is almost
coherent contrary to the magnetization curves calculated along $Ox$
where we find non uniform distributions of the magnetic moments at
the tips which act as nucleation points for the magnetization reversal.

\subsection{Infinite periodic arrays}

The chosen elementary cell corresponds to an array of 30 nanowires.
We have considered a finite number of $20\times20$ virtual copies
(12000 interacting nanowires) of the elementary cell \cite{nmag-hybrid}
for the calculation of the dipolar coupling between the nanowires.
The periodic boundary conditions have been applied for different inter-wire
distances: $a=15,$ 20, 30 and 40 nm. We have thus perform micromagnetic
simulations for hexagonal arrays of nanowires with a packing density
$P$ in the range \{$0\%-40\%$\}. In the case $a=15$ nm ($P\sim40\%$),
the periodic boundary conditions can-not be applied properly since
the separation between the nanowires is smaller than 2$d$ along the
$Oz$ direction. However, simulations can nevertheless be done by
considering small defects in the geometrical arrangement along the
$Oz$ direction (see supplementary Figure S2). \\

\begin{figure}[!h]
\includegraphics[bb=30bp 120bp 510bp 580bp,clip,width=8.5cm]{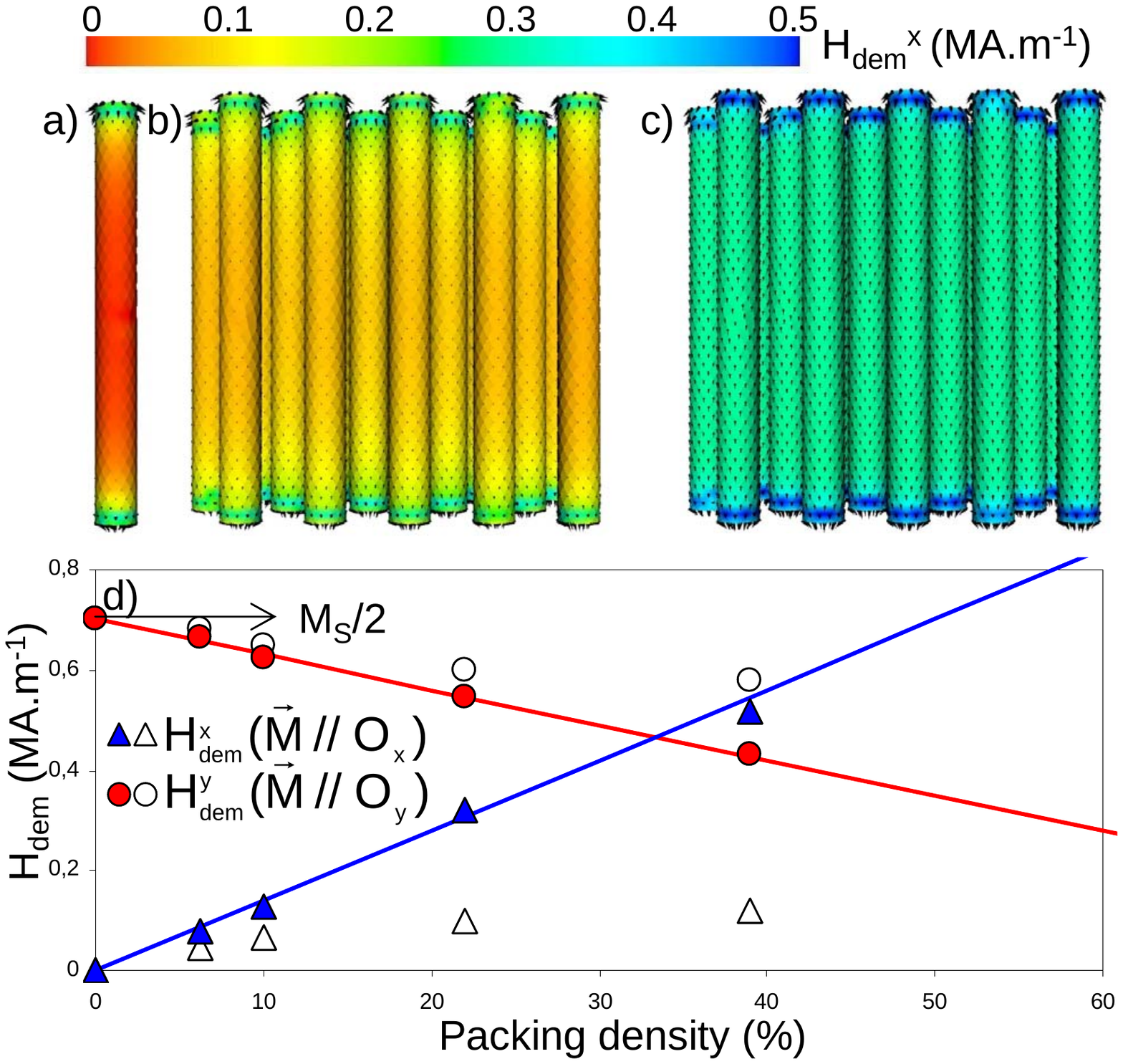}

\caption{a-b-c) Side view of the demagnetizing field distribution acting inside
the nanowire at remanence (after applying a saturating field along
$Ox$) for: a) an isolated nanowire, b) a finite array of 30 nanowires
with $a=20$ nm and c) an infinite array of nanowires (elementary
cell: 30 nanowires with $a=20$ nm). The colors encode the $x$-component
of the demagnetizing field $-H_{dem.}^{x}\cdot$ ($\times10{}^{6}$A.m$^{-1}$).
d) $x$- (filled triangles) and $y$- (filled circles) components
of the demagnetizing field measured in an infinite array of nanowires
homogeneously magnetized respectively along $Ox$ and $Oy$ as a function
of the packing density $P$. The solid lines correspond to the mean
field modelization \cite{delatorre,piraux}. Unfilled symbols correspond
to the values obtained in finite size arrays of 30 interacting nanowires. }

\end{figure}

For an applied magnetic field $B\parallel Ox$, for $a\geqslant20$
nm, the remanent states consist of wires magnetized in the same direction
along $Ox$ which is similar to what was obtained in the case of finite
arrays (Figure 5b and 6a) where the remanence was 1. However, in the
case $a=15$ nm, a fraction of wires ($\sim15\%$) have switched,
so that the remanence is only 0.7. This is due to the fact that for
large packing densities, the demagnetizing fields start to play a
significant role and are able to overcome the wires shape anisotropy.
This aspect is completely overlooked in finite size array simulations.
This is illustrated on Figure 9 where the hysteresis cycles in the
case of finite size and infinite size arrays calculations are compared.
Very large differences can be observed in the shapes of the hysteresis
curves. Figure 8 presents the demagnetizing field distributions inside
an isolated nanowire, a finite array of 30 nanowires ($a=20$ nm)
and an infinite array of nanowires (elementary cell of 30 nanowires
with $a=20$ nm). For an isolated nanowire, the demagnetizing field
$\vec{H}_{dem}$ is localized at the tips of the nanowire and is close
to zero in the central part of the nanowire \cite{ott2009,maurer2009}.
When 30 nanowires interact, $\vec{H}_{dem}$ is reinforced inside
the nanowires ($\sim0.1\times10^{6}$ A.m$^{-1}$) and takes a value
of about $0.3\times10^{6}$ A.m$^{-1}$ when an infinite array is
considered. Thus, as we shall see, the magnetization reversal will
be modified by the packing density $P$. These calculated values of
the demagnetizing field can be compared to the mean field model \cite{delatorre,piraux}.
Indeed, in a simple model, when all the nanowires are uniformly magnetized
along $Ox$, two limiting cases can be considered as a function of
$P$: i) $P=0\%$ corresponds to the case of an isolated nanowire
where we find that the $x$-component of the demagnetizing field is
close to zero, ii) $P=100\%$ corresponds to the continuous film limit
where $x$-component of the demagnetizing field is equal to $M_{S}$.
Our numerical results follow well this model (see Figure 9) and validate
our numerical calculations of the dipolar interactions. Note that
the mean field model can be applied only in the case of homogeneously
magnetized systems whereas our simulations can take into account the
dipolar coupling in the case of non uniform distributions of the magnetic
moments inside the nanowires.\\

\begin{figure}[!h]
\includegraphics[bb=39bp 2bp 470bp 595bp,clip,width=8.5cm]{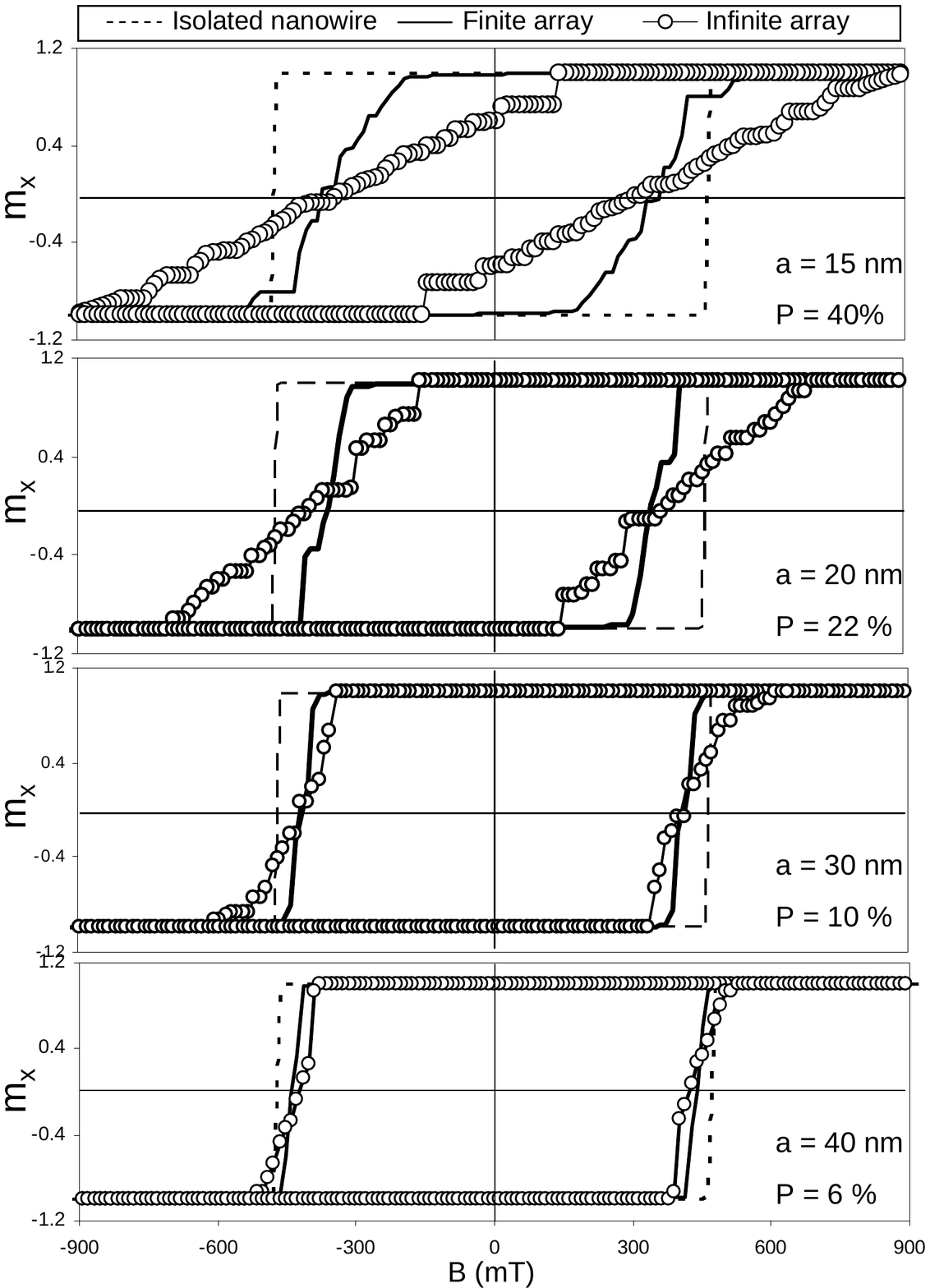}

\caption{$B\parallel Ox$: Hysteresis curves obtained from finite (30 nanowires:
solid line) and infinite (elementary cell of 30 nanowires: circles)
hexagonal arrays with different inter-wire distances ($a=15$, 20,
30, 40 nm). In each panel, the dashed lines represent the hysteresis
cycle for an isolated nanowire.}

\end{figure}

On the other hand, the magnetization curves calculated along $Ox$
(Figure 10) show strong changes, especially for large $P$ ($a\leq20$
nm) compared to the curves calculated with 30 interacting nanowires
(Figure 5). From Figure 10, we can show that, when decreasing the
inter-wire distance, the hysteresis curve calculated along $Ox$ is
less square and the saturation of the system (all magnetic moments
aligned along $Ox$ or $-Ox$) is reached for a larger applied field
\cite{nielsch}. Moreover, the remanent magnetization is reduced to
0.7 in the case of large $P$. For the larger inter-wire distance
($a=40$ nm), the hysteresis cycles do not strongly differs from the
one obtained in the finite array case but, when approaching the saturation,
the cycle departs from a square one. Figure 11 presents a top view
of the magnetic moments distribution during the magnetization reversal.
These micromagnetic configurations have been extracted from the curves
calculated with $P=22\%$. We show that the nanowires switch more
randomly compared to the case of finite arrays (see Figure 6).\\

\begin{figure}[!h]
\includegraphics[bb=30bp 250bp 550bp 585bp,clip,width=8.5cm]{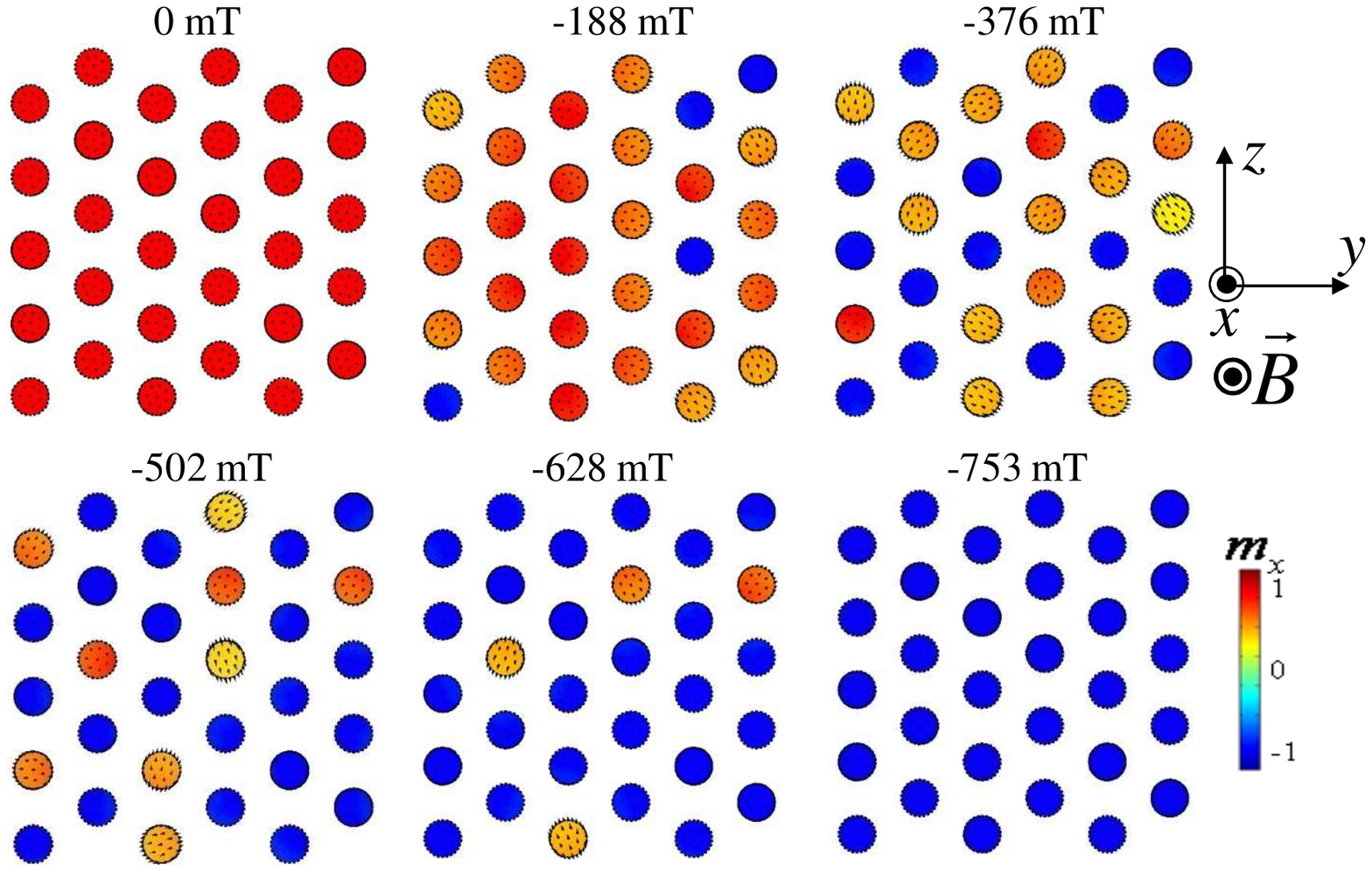}

\caption{Periodic array of nanowires (elementary cell $=30$ nanowires; $a=20$
nm): top view of the magnetic moments distribution as a function of
$B_{x}$. Colors encode $m_{x}$.}

\end{figure}

For $B\parallel Oy$ (or equivalently $B\parallel Oz$), the remanent
states do not differ from the ones obtained in finite arrays of 30
interacting nanowires: they correspond to a random orientations $+Ox$
or $-Ox$ of the nanowire magnetization. Moreover, the magnetization
reversal is similar to that observed in finite hexagonal array of
30 interacting nanowires, and a smaller $\mu_{0}H_{S}$ is necessary
to saturate the system as the inter-wire distance decreases (see Figure
12). The fact that magnetization curves calculated along $Oy$ (resp.
along $Ox$) tend to have smaller $\mu_{0}H_{S}$ (resp. $\mu_{0}H_{C}$)
when $a\rightarrow10$ nm or $P\rightarrow100\%$ (i. e. that the
hard (resp. easy) axis is less pronounced) is due to the continuous
film limit ($P=100\%$). In this limit (and in absence of magnetocrystalline
anisotropy), the $Ox$ direction is a hard axis whereas ($yOz$) becomes
an easy-plane for the magnetization \cite{thurn}. In addition, note
that when the system is saturated along $Oy$ (or $Oz$), the value
of the $y$-component of the demagnetizing field is well fitted by
using the mean field model \cite{delatorre,piraux} as presented in
Figure 9: we find that $H_{dem}^{y}=M_{S}/2$ for $P=0\%$ (isolated
nanowire) and $H_{dem.}^{y}$ linearly decreases and tends to zero
when $P$ increases (continuous film limit). Moreover, we find that
a crossing occurs around $P\sim1/3$ between the demagnetizing field
calculated along $O_{x}$ and $O_{y}$. A similar behaviour is observed
in the mean field model \cite{delatorre,piraux} and it corresponds
to the switching of the easy axis from the $O_{x}$ direction to the
plane ($yOz$).\\

\begin{figure}[!h]
\includegraphics[bb=42bp 380bp 382bp 585bp,clip,width=8.5cm]{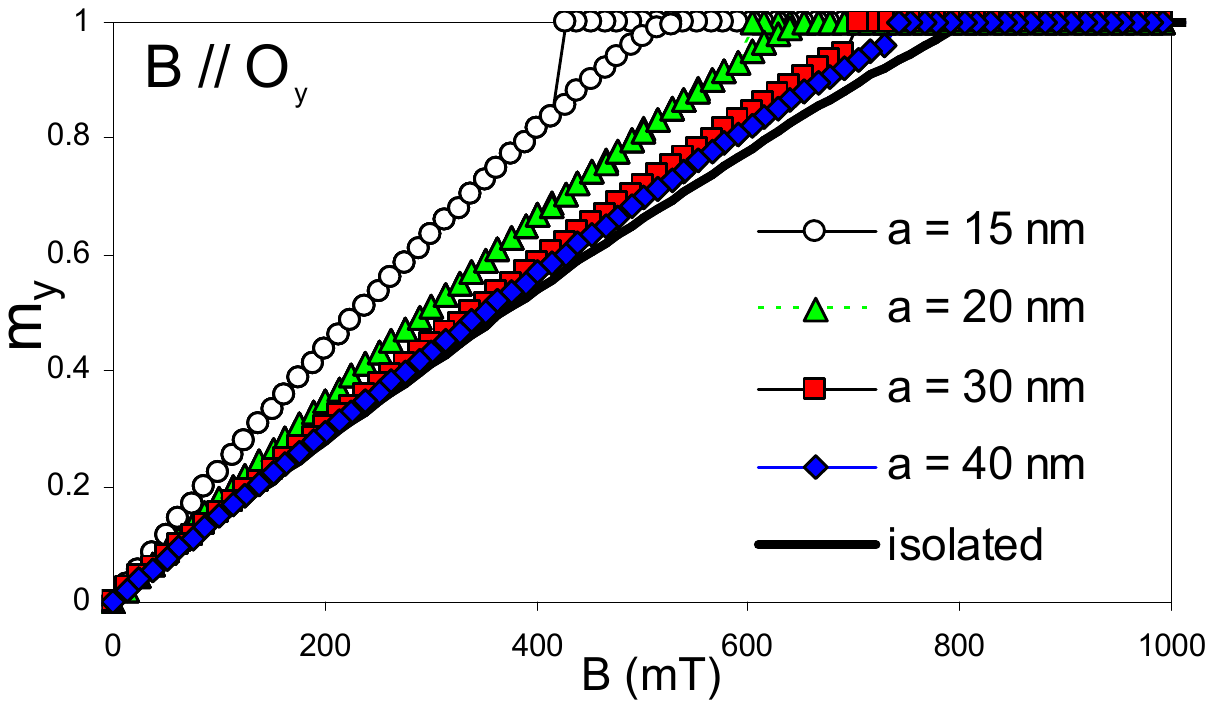}

\caption{$B\parallel Oy$ (or equivalently $B\parallel Oz$): magnetization
curves obtained from periodic (infinite) arrays with different inter-wire
distances ($a=15$, 20, 30, 40 nm).}

\end{figure}

These results show that infinite and dense hexagonal arrays of nanowires
need rigorous modelizations to take into account the dipolar coupling
whereas infinite arrays with larger inter-wire distance can be analyzed
by using a sufficiently large but finite number of interacting nanowires.\\

\section{Conclusion}

We have shown that user friendly tools such as the micromagnetic package
Nmag allow to perform calculation on complex extended systems. The
systematic study of more or less dense systems has demonstrated the
need to properly take into account the dipolar field. The use of finite
size models which do not properly include dipolar fields effects fail
to quantitatively reproduce the magnetic behavior of dense arrays
of nanowires with packing densities larger that $P\sim10\%$. We also
show that even for wires whose size is well below the radius of coherent
rotation, complex non homogeneous magnetic configurations develop
in the wires under applied magnetic field. These effects are overlooked
by effective field models but play a key role in the magnetic dynamics
properties of these wires. A natural extension of this work will be
to study the dynamics of the magnetization of these arrays and quantify
the effects of the non homogeneous magnetization and demagnetizing
field distributions inside the wires.\\

\begin{acknowledgments}
The authors gratefully acknowledge the Agence Nationale de la Recherche
(ANR) for their financial support (project 07-NANO-009 MAGAFIL). We
thank the NMAG developers for their advices.\end{acknowledgments}

\end{document}